# Signal duration and the time scale dependence of signal integration in biochemical pathways

Jason W. Locasale[1]

[1]Department of Biological Engineering, Massachusetts Institute of Technology, 77 Massachusetts Ave., Cambridge, MA 02139.

locasale@MIT.edu

**Abstract**

Signal duration (e.g. the time scales over which an active signaling intermediate persists) is a key regulator of biological decisions in myriad contexts such as cell growth, proliferation, and developmental lineage commitments. Accompanying differences in signal duration are numerous downstream biological processes that require multiple steps of biochemical regulation. Here, we present an analysis that investigates how simple biochemical motifs that involve multiple stages of regulation can be constructed to differentially process signals that persist at different time scales. We compute the dynamic gain within these networks and resulting power spectra to better understand how biochemical networks can integrate signals at different time scales. We identify topological features of these networks that allow for different frequency dependent signal processing properties. Our studies suggest design principles for why signal duration in connection with multiple steps of downstream regulation is a ubiquitous control motif in biochemical systems.

Signal duration (e.g. the length of time over which a signaling intermediate is active) is a critical determinant in mediating cell decisions in numerous biological processes(Fig. 1) including cell growth, proliferation, and developmental lineage commitments (Marshall 1995; Chen et al. 1996; Dolmetsch et al. 1997; Chen et al. 2001; Murphy et al. 2002; Sasagawa et al. 2005; Murphy and Blenis 2006; Santos et al. 2007). One fundamental issue in signal transduction and cell decision making then is how differences in signal duration are detected to achieve the appropriate biological response.

Accompanying changes in signal duration are multiple stages of biochemical regulation of differing network topology that collectively integrate an incoming signal to deliver a specific biological response. The sequential activation of multiple steps in a biochemical pathway is a ubiquitous regulatory motif involved in many aspects of gene regulation, metabolism, and intracellular signal transduction. Many advantages of having multiple steps of regulation as opposed to having activation occur through a single step have been documented. A signaling cascade can allow for attenuation of noise, incorporation of additional regulatory checkpoints or proofreading steps, and increased tunability of the input signal(Ferrell and Machleder 1998; Asthagiri and Lauffenburger 2001; Swain and Siggia 2002; Thattai and van Oudenaarden 2002). Other studies have also established conditions under which signaling cascades amplify or attenuate incoming signals(Heinrich et al. 2002; Chaves et al. 2004; Locasale et al. 2007). These conditions are established by rates of activation, rates of deactivation that are set by phosphatase activities, and the presence of scaffold proteins. However, how these features synergize to detect differences in signal duration has not been fully studied.

A recent study has proposed a model that predicts how signals with different dynamical characteristics can be distinguished upon integration into network architectures(Behar et al. 2007a; Behar et al. 2007b). We develop a formalism to complement their approach and, as a consequence, identify general principles for how different network topologies can differentially integrate signals that persist at different time scales. We focus on a simple model of the sequential enzymatic activation of multiple species along a pathway to understand mechanistic principles underlying how multiple stages in a biochemical pathway can integrate differences in signal duration. We use a model of a weakly activated cascade(Heinrich et al. 2002; Chaves et al. 2004; Bardwell et al. 2007), whose assumptions we first motivate, to investigate general principles underlying how biochemical cascades detect the time scale dependence of input signals. The model allows us to characterize the dynamics by obtaining exact expressions for the power spectra of biochemical networks of multiple stages with arbitrary length and connectivity.

Our analysis emphasizes how signals that persist at different time scales multiple staged biochemical pathway. We first show that biochemical cascades can function as both high low and high pass filters depending on the topology of the network architecture. A low pass filter removes high frequency (short duration) components of a signal and a high pass filter removes low frequency (long duration) components of a signal. These filtering capabilities are determined by differential positive and negative regulation within the biochemical module that occurs at different time scales. Importantly, the filtering capabilities are determined by the presence of feedback as well as the amplification and attenuation properties at different steps in the cascade that are set by the differences in phosphatase activities at different stages along the cascade. Our findings suggest design principles that characterize how biochemical cascades are well suited for detecting time scale dependent differences in biochemical signals.

An input signal, $f(t)$, activates the first member of the pathway whose activation can then activate the next member. In turn, each upstream species activates its immediate downstream target and can also be deactivated by, for example, a phosphatase. Assuming Michaelis Menten kinetics for the activation and deactivation of each species along the cascade, we can write an equation for the dynamics of the active form of the $i^{th}$ species $x_i$ along the cascade:

$$\frac{dx_i}{dt} = \frac{k_{cat}^i x_{i-1}\left(x_i^T - x_i\right)}{K_M^i + \left(x_i^T - x_i\right)} - \frac{k_{cat,pase}^i E_{pase} x_i}{K_{M,pase}^i + x_i}, \qquad (1)$$

where $E_{pase}$ is the concentration of the enzyme that deactivates species $i$, $k_{cat}^i, k_{cat,pase}^i$ are the catalytic constants of the activation and deactivation steps, $k_M^i, k_{M,pase}^i$ are the Michaelis constants and $x_i^T$ is the total amount species available at step i. We take $K_M^i, K_{M,pase}^i$ at each step to be large ($K_M^i, K_{M,pase}^i > x_i^T$) so that the kinetics of the reactions are not limited by the availability of the enzyme(Goldbeter and Koshland 1981). Next, if we assume that the cascade is weakly activated (i.e. at each stage, the total number of species is much larger than the number of active species, $x_i^T \gg x_i$). In many biologically relevant instances (e.g. the Mitogen activated protein kinase (MAPK) cascade), the neglect of saturation effects is reasonable(Hornberg et al. 2005). Further, modeling the deactivation as a first order reaction is often valid when phosphatases are in excess as is the case in many physiological scenarios(Ghaemmaghami et al. 2003). Eq. 1 simplifies to a system of linear first order differential equations:

$$\frac{dx_1}{dt} = k_1^+ f(t) - k_1^- x_1$$
$$\frac{dx_i}{dt} = k_i^+ x_{i-1} - k_i^- x_i \quad , \quad (2)$$

where the first species is activated at a rate $k_1^+ f(t)$; $k_i^+ = \frac{x_i^T k_{cat}^i}{K_{M,i}}$ and $k_i^+ = \frac{E_{pase} k_{cat,pase}^i}{K_{M,pase}^i}$.

The weakly activated cascade model has the advantage that it is analytically tractable. Eq. 2 can be conveniently analyzed by introducing Fourier transformed variables: $X_i(\omega) = \int dt e^{i\omega t} x_i(t)$ and $F(\omega) = \int dt e^{i\omega t} f(t)$. The number of activated species at stage $i$ becomes $X_i(\omega) = \frac{k_i^+ X_{i-1}(\omega)}{i\omega + k_i^-}$. The power spectrum $P_i(\omega) \equiv |X_i(\omega)|^2$ at the $i^{th}$ step can be obtained $P_i(\omega) = \frac{(k_i^+)^2}{\omega^2 + (k_i^-)^2} P_{i-1}(\omega)$. After iterating at each successive stage of the $n$ step cascade, an expression for $P_n(\omega)$ as a function of the power spectrum of the input signal ($S(\omega) \equiv |F(\omega)|^2$) is obtained:

$$P_n(\omega) = g_n(\omega) S(\omega), \quad (3)$$

in which a frequency dependent gain $g_n(\omega)$ is defined as:

$$g_n(\omega) = \prod_{i=1}^{n} \left(\frac{k_i^+}{k_i^-}\right)^2 \frac{1}{1 + \left(\frac{\omega}{k_i^-}\right)^2} . \quad (4)$$

$g_n(\omega)$ is a transfer function that converts the input $S(\omega)$ into a response and provides a measure of the signal processing capabilities of the network. The change in the amplitude of the signal output is determined by the $\prod_{i=1}^{n}\left(\frac{k_i^+}{k_i^-}\right)^2$ term and the time scale dependence of the output is modulated by the $\prod_{i=1}^{n}\left[1+\left(\frac{\omega}{k_i^-}\right)^2\right]^{-1}$ term.

From the formula of $g_n(\omega)$, one consequence of having multiple stages is readily apparent. In the high frequency regime $\left(\frac{\omega}{k_i^-} \gg 1\right)$, for each $k_i^-$, $g_n(\omega)$ rapidly decays with increasing $n$ ($g_n(\omega) \sim \omega^{-2n}$). Thus, longer cascades are more efficient at filtering the high frequency components of the signal from the output. This behavior is illustrated in Fig. 2b. Fig. 2b contains plots of $g_n(\omega)$ for different values of $n$; cascades of lengths $n = 1, 2, 3, 4$ are shown.

In eq. 4, the relative values of $k_i^-$ along different stages of the cascade also affects the scaling behavior of $g_n(\omega)$ as $\omega$ changes as well as the overall amplitude. The change in signal amplitude at the steady state (that leads to amplification or attenuation) at step $i$ is given by the ratio of the effective rate constants for activation and deactivation $\frac{k_i^+}{k_i^-}$. $\frac{k_i^+}{k_i^-} > 1$ results in signal amplification and $\frac{k_i^+}{k_i^-} < 1$ leads to attenuation of the signal at step $i$ (Heinrich et al. 2002; Chaves et al. 2004). Amplification or attenuation also leads to different frequency dependent behaviors of $g_n(\omega)$. For example, consider an n staged cascade with rate constants $k_i^-$ such that $k_n^- > k_{n-1}^- > k_{n-2}^-...k_2^- > k_1^-$. At frequencies $\omega > k_n^-$, $g_n(\omega) \sim \omega^{-2n}$ while at frequencies $k_n^- > \omega > k_{n-1}^-$, $g_n(\omega) \sim \omega^{-2(n-1)}$ and so forth. Thus, at intermediate frequencies, a time scale separation (as determined by different deactivation rates along the cascade), along with signal amplification and attenuation, also leads to different frequency dependent behaviors of $g_n(\omega)$.

Fig. 2c contains plots of $g_n(\omega)$ for $n = 1, 2, 3, 4$ with successively different values of $k_i^-$ while keeping $\dfrac{k_i^+}{k_i^-} = 1.0$ fixed ($k_1^- = 1.0$, $k_2^- = 3.3$, $k_3^- = 6.6$, $k_4^- = 10.0$). Also, from the plots in Fig. 2c (and inspection of eq. 4), it is observed that incorporation of faster steps along the cascade influences the frequency dependence of $g_n(\omega)$ less than would be case when kinetics of activation are same for each successive step. The faster steps are effectively removed from $g_n(\omega)$. This observation suggests a design principle in the ability of a biochemical pathway to filter signals of short duration: when there is a positive gradient of deactivation rates (that also leads to amplification or attenuation of signal amplitude), the time scale dependence of signal integration for multi staged cascades more closely resembles that of a pathway involving a single step.

While the sequential activation of multiple steps in a biochemical pathway allows for effective filtering of the high frequency components of a signal, often the desired signal output is regulated by feedback. We will now show that feedback control in some instances also allows for the filtering of the low frequency (long time) components of an active signal. In these instances, signals that occur at short times can be integrated while signals with a longer duration are effectively filtered.

For instance, the signal output can be affected by a feedback loop that is initiated downstream of the output. This scenario would be the case when the signal output from a biochemical cascade activates its own positive or negative regulators. For instance, in mammalian cells, the activation of extra cellular regulatory kinase (ERK) often leads to the upregulation or activation of its own phosphatases(Amit et al. 2007). In this scenario, signal output is activity of the kinase at the $m^{th}$ step and feedback control to the signal output at step $m$ is initiated at a later step (i.e. $n > m$) then the modified set of dynamical equations becomes:

$$\frac{dx_1}{dt} = k_1^+ f(t) - k_1^- x_1$$
$$\frac{dx_m}{dt} = k_m^+ x_{m-1} - k_m^- x_m + \upsilon \, k^f x_n \tag{5}$$

where $k^f$ is the feedback strength and sets the time scale of the feedback and
$\upsilon = \begin{cases} 1 \; ; positive \; feedback \\ -1 \; ; negative \; feedback \end{cases}$. After applying a Fourier transformation as before, an expression for $P_m^f(\omega)$, albeit now more complicated, can be obtained as a function of $S(\omega)$ in closed-form:

$$P_m^f(\omega) = g_m^f(\omega) S(\omega), \tag{6}$$

where,

$$g_m^f(\omega) = \prod_{i=1}^{m} \frac{\left(k_i^+\right)^2}{\left(\left(k_i^-\right)^2 + \omega^2\right)}$$

$$+ \frac{\left(k^f\right)^2 \left[\prod_{i=1}^{n}\left(k_i^+\right)^2\right]}{\left(\left(k_m^-\right)^2 + \omega^2\right)\prod_{i=1}^{n}\left(\left(k_i^-\right)^2 + \omega^2\right) + \left(k^f\right)^2 \prod_{i=m+1}^{n}\left(k_i^+\right)^2 \prod_{i=1}^{m-1}\left(\left(k_i^-\right)^2 + \omega^2\right) + 2\prod_{i=m}^{n}\left(k_i^+\right)^2 \prod_{i=1}^{m-1}\left(\left(k_i^-\right)^2 + \omega^2\right)}$$

$$+ \frac{\upsilon}{2} \frac{k^f \left[\prod_{i=1}^{n}\left(k_i^+\right)\right]\left[\prod_{i=1}^{m}\left(k_i^+\right)\right]}{\left[\prod_{i=m}^{n}\left(k_i^-\right)\right]\prod_{i=1}^{m}\left(\left(k_i^-\right)^2 + \omega^2\right) - \upsilon k^f \left(\left(k_m^-\right)^2 + \omega^2\right)\prod_{i=m+1}^{n}\left(k_i^+\right)\prod_{i=1}^{m-1}\left(\left(k_i^-\right)^2 + \omega^2\right)}$$

(7).

In the case of feedback regulation, there exists a competition between processes that are realized on multiple time scales: one for the signal to propagate along the cascade to the species involved in the signal output, and the others for the additional interactionss derived from the feedback loops to propagate and interact with the species involved in the output. The competition between these effects in principle may lead to a frequency dependent optimal value of $g_m^f(\omega)$. At high frequencies as before, signal propagation is limited by the time it takes to move through the cascade and high frequency components of $g_m^f(\omega)$ are filtered. Also, at low frequencies, signals can potentially be attenuated when the response is dominated by the activity of the feedback loop. If the feedback loops is sufficiently strong, then the low frequency components of the signal are also filtered by the cascade. In this scenario, the behavior of $g_m^f(\omega)$ would non monotonic.

We illustrate these ideas through consideration of a three tiered cascade that consists of a chemical species carrying the input signal, a species conferring the signal output, and a species activated downstream to the output that provides a feedback interaction to the species conferring the signal output. In this scheme, $m = 1$ and $n = 2$, and eq. 7 is simplified and $g_1^f(\omega)$ becomes:

$$g_1^f(\omega) = \frac{(k_1^+)^2 \left[(k_2^-)^2 + \omega^2\right]}{(k_1^- k_2^- - \upsilon k_f k_2^+)^2 + \omega^2 \left((k_1^-)^2 + (k_2^-)^2 + 2\upsilon k_f k_2^+ + \omega^2\right)}. \tag{8}$$

The optimal frequency $\omega_{opt}$ is obtained by differentiating $g^f{}_1(\omega)$,

$$\omega_{opt} = \left[\left\{k_f k_2^+ \left(k_f k_2^+ - 2\upsilon k_2^- (k_1^- + k_2^-)\right)\right\}^{1/2} - (k_2^-)^2\right]^{1/2}. \tag{9}$$

$\omega_{opt}$ increases monotonically for decreasing values of $k_2^-$ and increasing values of $k_f$. For negative feedback $\upsilon = -1$, $\omega_{opt}$ exists ($\text{Im}\,\omega_{opt} = 0$) when

$\left[k_f k_2^+ \left\{k_f k_2^+ + 2k_2^- (k_1^- + k_2^-)\right\}\right]^{1/4} > k_2^-$. Positive feedback $\upsilon = +1$ requires that two conditions are satisfied for $\omega_{opt}$ to be real, $\left[k_f k_2^+ \left(k_f k_2^+ - 2k_1^- k_2^-\right)\right]^{1/4} > k_2^-$ and

$k_f k_2^+ > 2k_2^- (k_1^- + k_2^-)$. The height at the optimal frequency $g_1^f(\omega_{opt})$ is:

$$g_1^f(\omega_{opt}) = \frac{(k_1^+)^2}{(k_1^-)^2 - (k_2^-)^2 + 2\left\{k_f k_2^+ + \sqrt{k_f k_2^+ \left[k_f k_2^+ - 2k_2^- (k_1^- + k_2^-)\right]}\right\}}. \tag{10}$$

We can also compute the width $\omega_{1/2}$ of $g_1^f(\omega)$ at half maximum $g_1^f(\omega_{opt})/2$. $\omega_{1/2}$ has the form: $\omega_{1/2} = \frac{1}{\sqrt{2}}\left[\sqrt{4\gamma_1 + \gamma_2 + \sqrt{8\gamma_1(2\gamma_1 + 2\gamma_3) + \gamma_4}} - \sqrt{4\gamma_1 + \gamma_2 - \sqrt{8\gamma_1(2\gamma_1 + 2\gamma_3) + \gamma_4}}\right]$,

(11)

where $\gamma_1 = \sqrt{k_f k_2^+ \left[k_f k_2^+ - 2k_2^- (k_1^- + k_2^-)\right]}$, $\gamma_2 = (k_1^-)^2 - 3(k_2^-)^2 + 2k_f k_2^+$, $\gamma_3 = \gamma_2 + 2(k_2^-)^2$,

and $\gamma_4 = (k_1^- + k_2^-)^2 (k_1^- - k_2^-)^2 + 4k_f k_2^+$. Fig. 3b considers plots of $g_1^f(\omega)$ for different feedback strengths $k_f$. The curves in Fig. 3b illustrate changes in $g_1^f(\omega)$, $g_1^f(\omega_{opt})$, and $\omega_{1/2}$. Fig. 3c illustrates how $g_1^f(\omega)$, $g_1^f(\omega_{opt})$, and $\omega_{1/2}$ change for different values of $k_2^-$.

A convenient way to parameterize signals of differing duration, while keeping the total amount of signal $\int f(t)dt = \alpha$ fixed, is to consider the function

$$f(t) = \alpha \frac{te^{-t/\tau_d}\Theta(t)}{\int te^{-t/\tau_d}\Theta(t)dt},$$ where $\Theta(t)$ is a Heaviside step function, $\tau_d$ sets the signal duration, and $\alpha$ is taken to be 1 ($\alpha = 1$) in the appropriate units. This form of $f(t)$ models the behavior of a typical experimental signaling time course(Murphy and Blenis 2006). For this choice of signal, $S(\omega)$ is easily computed; $S(\omega) = \frac{\alpha^2}{\left(\tau_d^{-2} + \omega^2\right)^2}$. $S(\omega)$ is plotted in Fig. 4a for different values of $\tau_d$ ranging from $\tau_d^{-1} = 2.0$ (short duration) to $\tau_d^{-1} = 0.1$ (long duration). The corresponding plots of $f(t)$ are shown on the inset of Fig. 4a.

Figs. 4b,c illustrates how signals $S(\omega)$ of large ($\tau_d^{-1} = 0.5$ dotted lines) and small ($\tau_d^{-1} = 2.0$ dashed lines) duration are integrated by the internal gains $g_3(\omega)$ and $g_1^f(\omega)$ of these multistage cascades of differing network topologies. In Fig. 4b, the signal output $P_3(\omega)$, upon integration by a three-tiered kinase cascade is shown. Taking $k_i^+ = k_i^- = 1.0$ for $i \in 1,2,3$, $g_3(\omega)$ effectively filters the short ($\tau_d^{-1} = 2.0$) duration signal and results in an output $P_3(\omega)$ of small magnitude at all time scales $2\pi\omega^{-1}$ in the frequency spectrum. In contrast, for the signal characterized by $\tau_d^{-1} = 0.5$, signal processing through $g_3(\omega)$ results in a signal of larger amplitude. The ratio of amplitudes $\frac{P_3^{\tau_d=0.5}}{P_3^{\tau_d=2.0}}$ at the optimal frequency mode ($\omega = 0$) for the two signals is $\frac{P_3^{\tau_d=0.5}}{P_3^{\tau_d=2.0}} \approx 17$ In Fig. 4c, the signal output $P_1^f(\omega)$, obtained from a signal output that is also affected by a downstream negative ($\upsilon = -1$) feedback loop, is shown. Parameters used are: $k_1^+ = 2.0$, $k_2^+ = 1.0$, $k_1^- = 1.0$, $k_2^- = 0.01$, $k^f = -5.0$. For the signal of long duration $\tau_d^{-1} = 0.5$, only the small frequency components of the signal are integrated. This behavior is in contrast with the signal output of a short duration signal $\tau_d^{-1} = 2.0$. The amplitude difference in this case is $\frac{P_1^{f,\tau_d=0.5}}{P_1^{f,\tau_d=2.0}} \approx 0.2$.

We demonstrated how simple network structures involving multiple steps of biochemical regulation can differentially detect signals that have different temporal behaviors. Our models illustrate the topological requirements needed for different network motifs to detect signals of differing duration. To illustrate these effects, we computed the frequency dependent internal gain for two classes of biochemical pathways involving multiple stages of regulation. The first model consisted of a cascade of steps and showed how changes in the number of steps as well as the amplification/attention changed the networks' ability to filter high frequency (short duration) components of a signal. The second model consisted of a sequence of steps in which the output is connected to a downstream feedback loop. The gain in this network can have non monotonic behavior in which the low frequency components of the signal are also filtered at time scales commensurate with the induction of the regulatory loop. This behavior enables the network to filter out signals of long duration. The minimal topological features of these biochemical networks provide distinct and robust mechanisms for integrating signals that persist with different characteristic time scales.

**Fig. 1. Physiological examples of signal duration determining the phenotypic outcome in signal transduction**

Four examples of branching physiological processes in which phenotypic decisions are believed to be controlled by the detection of differences in signal duration(Marshall 1995; Murphy et al. 2002; Kortum et al. 2005; Daniels et al. 2006; El Kasmi et al. 2006; Santos et al. 2007). .

**Fig. 2. Filtering of high frequency signals**

Time scale dependence of signal integration in a biochemical cascade. a.) the sequential activation of multiple stages in a signaling cascade. Superscripts (I) and (A) denote inactive and active forms of each chemical species and are dropped from the equations in the text. b.) same kinetic constants, all kinetic constants are taken to be: $k_i^+ = k_i^- = 1.0$ c.) a positive gradient activation/deactivation rates keeping $\dfrac{k_i^+}{k_i^-} = 1.0$ fixed. $k_1^+ = 1.0$, $k_2^+ = 3.3$, $k_3^+ = 6.6$, $k_4^+ = 10.0$

**Fig. 3.) Filtering of low frequency signals**

A three tiered biochemical cascade with competing processes occurring at different time scales is sufficient to filter signals at long time scales (Low pass filtering). a.) An initial stimulus activates a downstream species that confers a signal output and activates a downstream species that, through feedback, interacts with the species that carries the signal output. Superscripts (I) and (A) denote inactive and active forms of each chemical species and are dropped from the equations in the text. b.) For each plot, these parameter values were taken to be: $\upsilon = -1$, $k_1^+ = k_1^- = k_2^+ = 1.0$ ; $k_2^- = 0.1$. $k_f = 2.5$ (solid lines), $k_f = 1.0$ (dashed lines), $k_f = 0.5$ (dotted lines), $k_f = 0.1$ (dash-dotted lines). c.) parameter values were taken to be: $\upsilon = -1$, $k_1^+ = k_1^- = k_2^+ = k_f = 1.0$. $k_2^- = 0.0$ (solid lines), $k_2^- = 0.5$ (dashed lines), $k_2^- = 1.0$ (dotted lines), $k_2^- = 2.0$ (dash-dotted lines).

**Fig 4.) Integration of differences in signal duration**

a.) Differences in signal duration parameterized by $\tau_{deg}$; $\tau_{deg}^{-1} = 0.1$ (dash-dotted), 0.5 (dotted), 1.0 (dashed), 2.0 (solid) lines. Plots of $S(\omega) \equiv |F(\omega)|^2$ are shown. Corresponding plots of $f(t)$ are shown in the inset. Short $\tau_{deg}^{-1} = 2.0$ and long $\tau_{deg}^{-1} = 0.5$ duration signals are filtered through b.) $g_3(\omega)$ and c.) $g_1^f(\omega)$ resulting in b.) $P_3(\omega)$ and c.) $P_1^f(\omega)$ for $\tau_{deg}^{-1} = 2.0$ (dashed lines) and $\tau_{deg}^{-1} = 0.5$ (dotted lines).

Parameters taken to be: b.) c) $k_1^+ = 2.0$, $k_2^+ = 1.0$, $k_1^- = 1.0$, $k_2^- = 0.01$, $k^f = -5.0$.

Fig. 1.

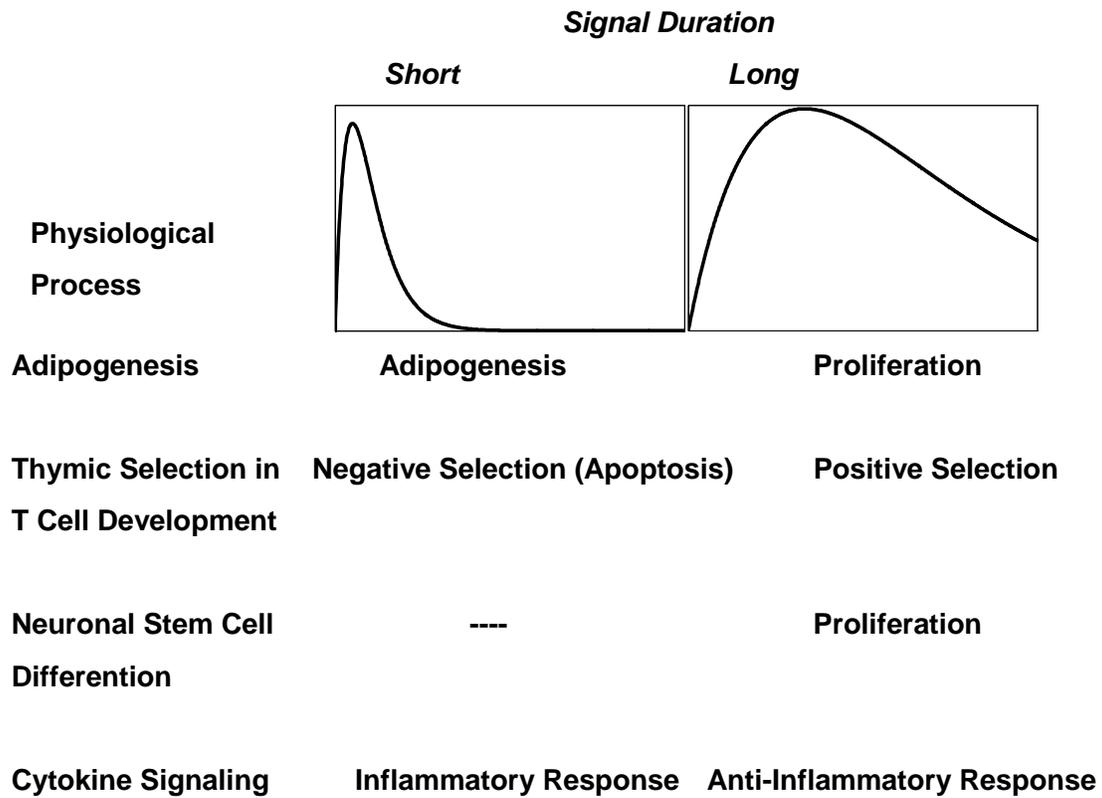

| Physiological Process | Short | Long |
|---|---|---|
| Adipogenesis | Adipogenesis | Proliferation |
| Thymic Selection in T Cell Development | Negative Selection (Apoptosis) | Positive Selection |
| Neuronal Stem Cell Differention | ---- | Proliferation |
| Cytokine Signaling | Inflammatory Response | Anti-Inflammatory Response |

*Signal Duration*

Fig. 2.

a.)

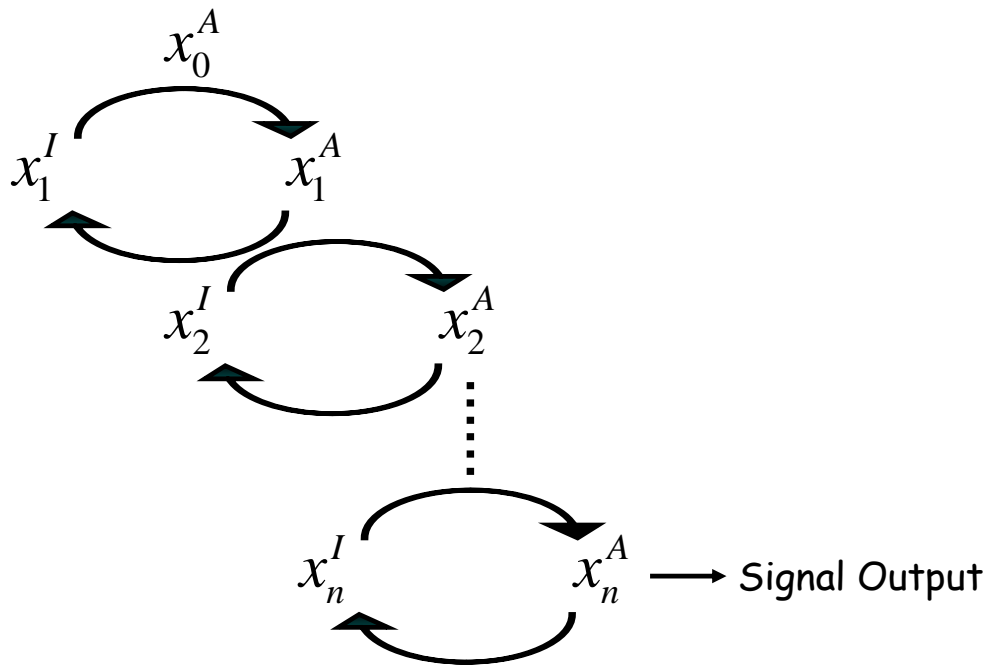

b.)

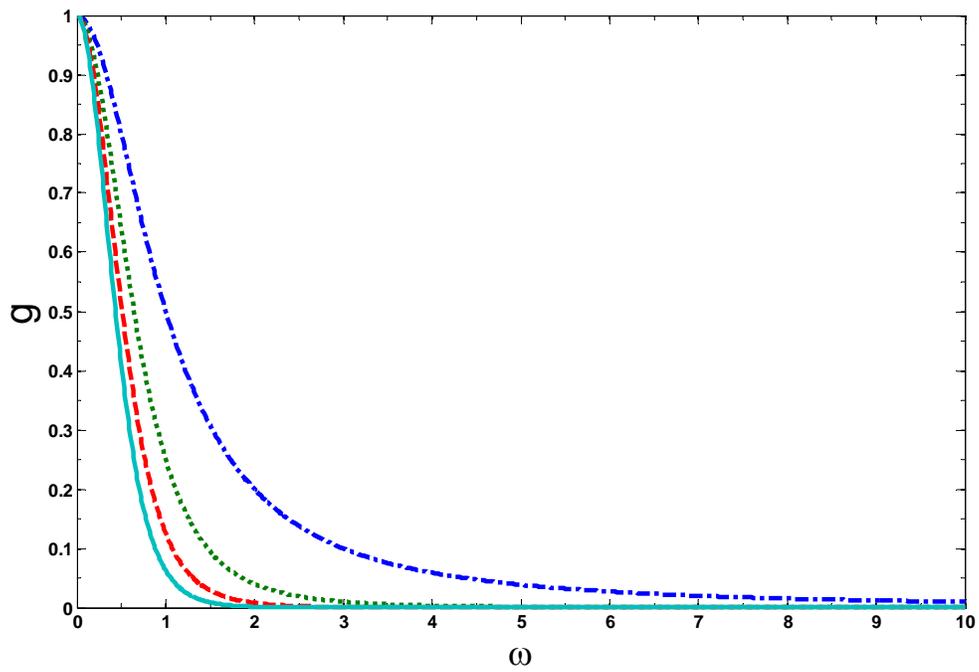

c.)

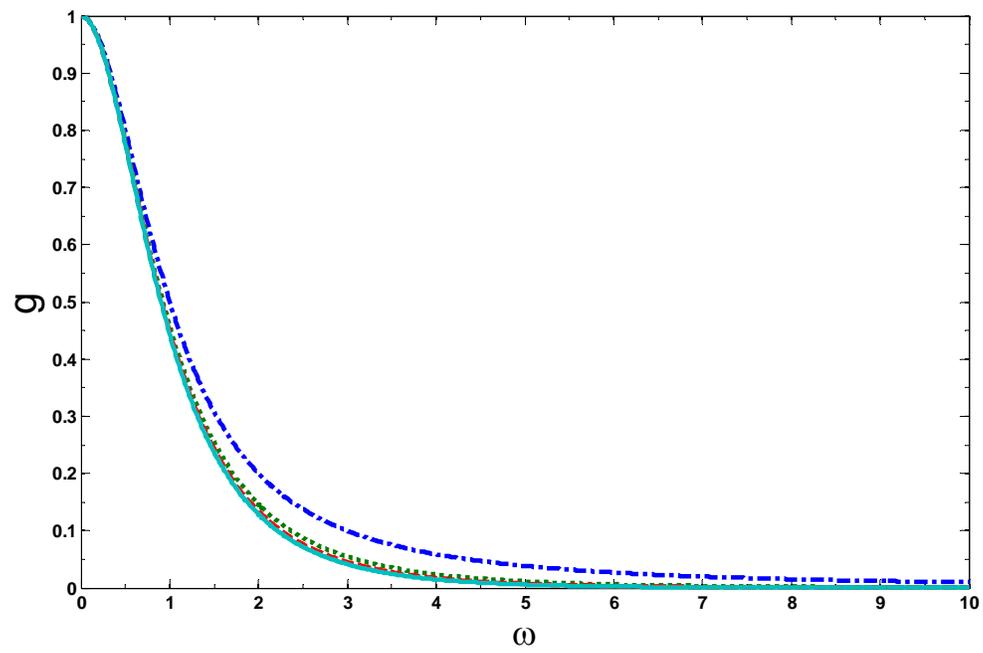

Fig. 3.

a.)

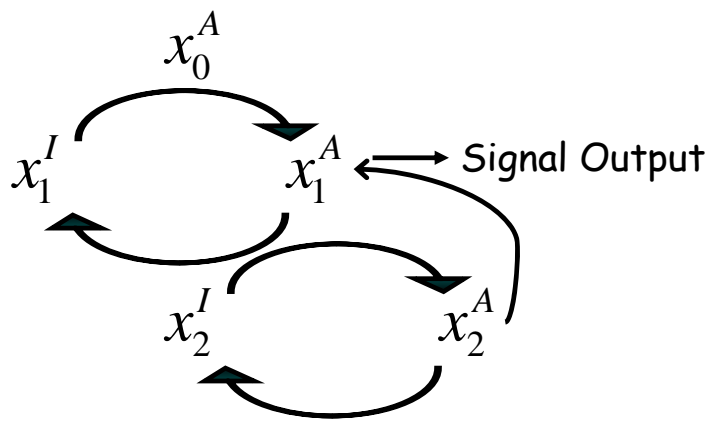

b.)

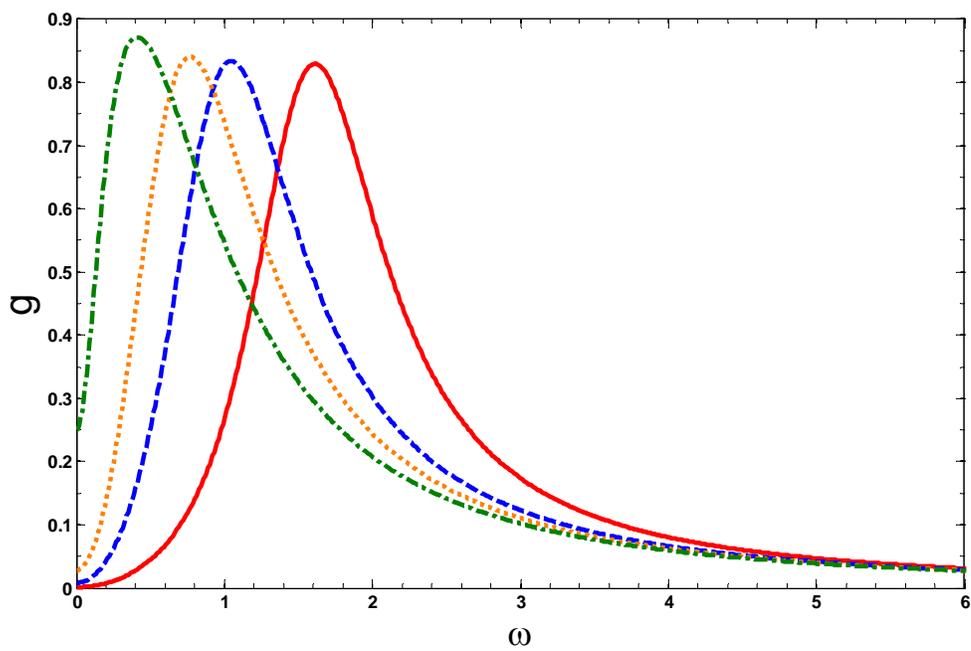

c.)

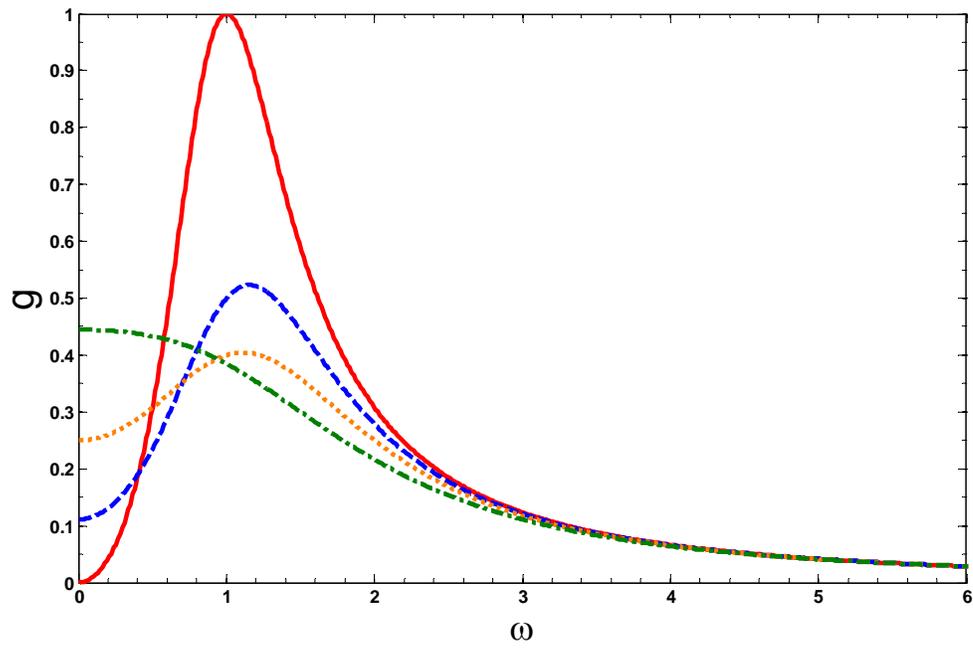

Fig. 4.

a.)

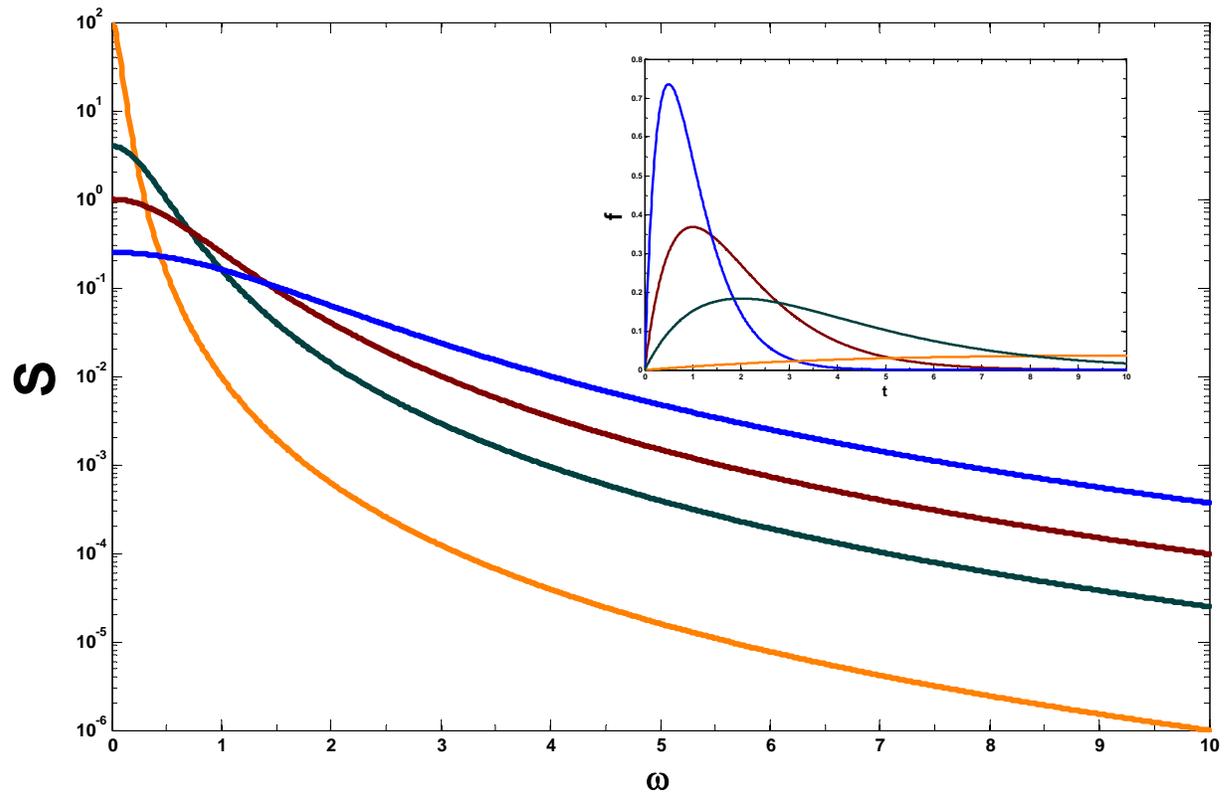

b.)

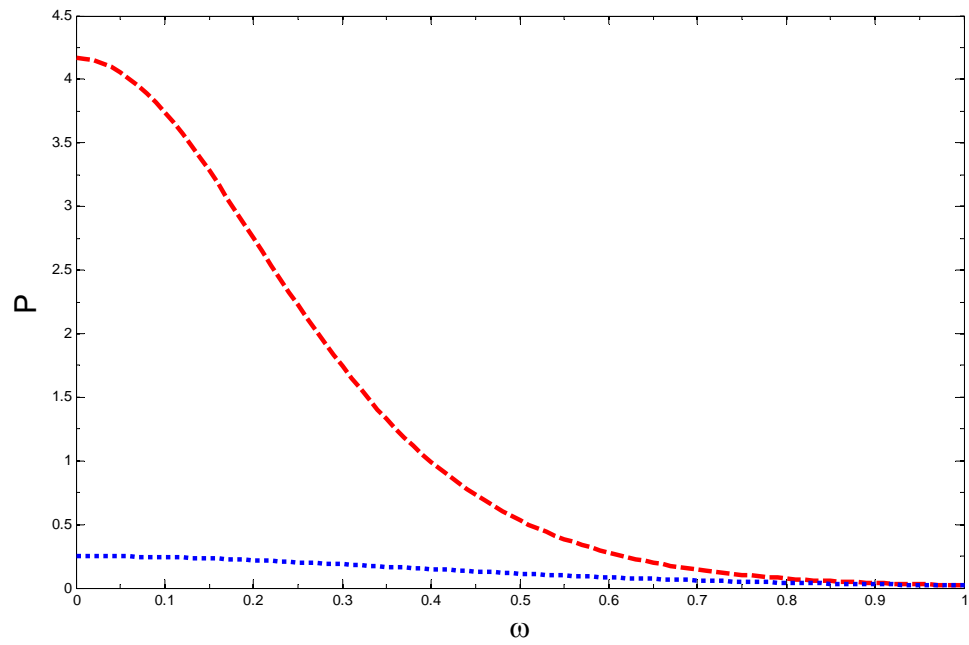

c.)

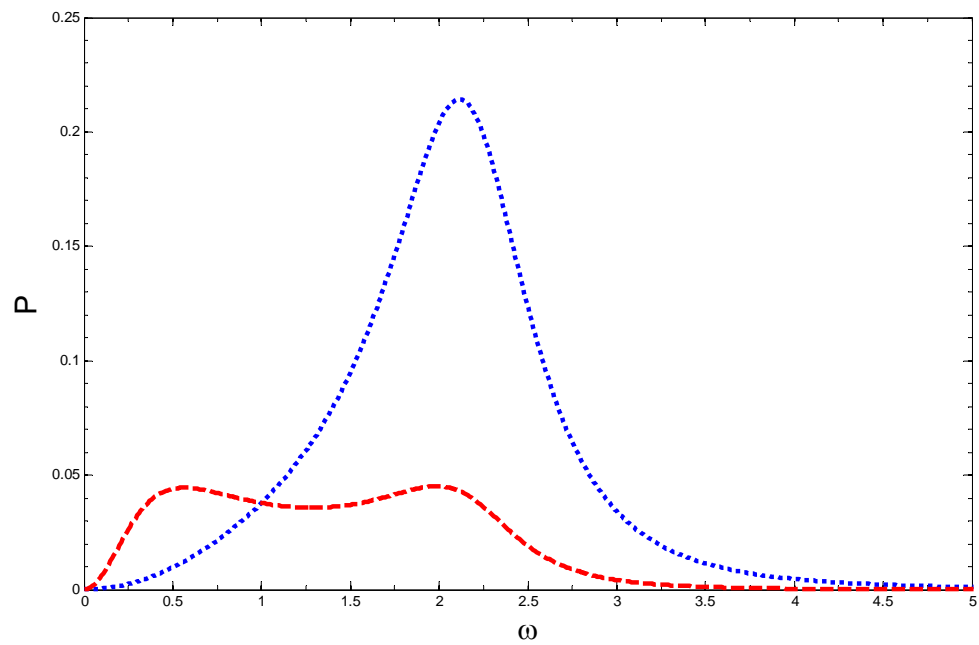